\documentclass[prd,eqsecnum,apsi,twocolumn]{revtex4}

\usepackage{graphicx}
\usepackage{latexsym}
\usepackage{amsmath}
\usepackage{amssymb}

\newcommand{\be}{\begin{equation}}
\newcommand{\ee}{\end{equation}}
\newcommand{\bea}{\begin{eqnarray}}
\newcommand{\eea}{\end{eqnarray}}

\begin{document}

\begin{flushleft} 
PITHA 09/32 \newline
RECAPP-HRI-2009-021
\end{flushleft}

\title{\Large \bf Bulk Higgs field in a Randall-Sundrum model with
  nonvanishing brane cosmological constant}

\author{Paramita Dey${}^{}$ \footnote{E-mail:
    paramita@physik.rwth-aachen.de}} \affiliation{Institut f\"ur
  Theoretische Physik E,\\ RWTH Aachen, D-52056 Aachen, Germany}

\author{Biswarup Mukhopadhyaya${}^{}$
  \footnote{E-mail: biswarup@mri.ernet.in}}
\affiliation{Regional Center for Accelerator-based Particle Physics,\\
  Harish-Chandra Research Institute, Chhatnag Road, Jhusi, Allahabad -
  211 019, India}

\author{Soumitra SenGupta${}^{}$ \footnote{E-mail: tpssg@iacs.res.in}}
\affiliation{ Department of Theoretical Physics and Center for Theoretical
Sciences,\\ Indian Association for the Cultivation of Science, Kolkata - 700
032, India}

\begin{abstract}
  We consider the possibility of Higgs mechanism in the bulk in a
  generalised Randall-Sundrum model, where a nonvanishing cosmological
  constant is induced on the visible brane. This scenario has the
  advantage of accommodating positive tension of the visible brane and
  thus ensures stability of the model. It is shown that several
  problems usually associated with this mechanism are avoided if some
  dimensionful parameters in the bulk are allowed to lie {\it a little
    below} the Planck mass. The most important of these is keeping the
  lowest massive mode in the scale of the standard electroweak model,
  and at the same time reducing the gauge coupling of the next excited
  state, thus ameliorating otherwise stringent phenomenological
  constraints.
\end{abstract}

\maketitle

\section{Introduction}
\label{secone}
Extra dimensional models with a warped background, first proposed by
Randall and Sundrum (RS) \cite{rs}, offer a novel explanation of the
hierarchy between the Planck ($M_P$) and electroweak (EW) energy
scales. Such theories postulate that our universe is five-dimensional,
described by the metric
\be 
\label{metric}
ds^2 = e^{-2\sigma (y)} \eta_{\mu\nu} dx^\mu dx^\nu - dy^2, 
\ee 
where $y~=~r_c\phi$ denotes the extra spacelike dimension compactified on
$S^1/Z_2$ with radius $r_c$, $\sigma (y) = ky$, and $k \equiv
\sqrt{-\Lambda/12M^3_P}$ denotes the curvature scale determined by the
negative 5-d cosmological constant $\Lambda$, and thus, is of the order of the
Planck mass ($M_P$). Two 3-branes are located at the boundaries at $y~=~0$ and
$\pi r_c$; the latter contains the physics of the standard model (SM) of
elementary particles and is called the `visible' brane. The exponential `warp'
factor provides the aforementioned hierarchy of mass parameters, once
projections on the `visible' brane are taken, for $kr_c~\simeq~12$
\cite{rs}. Since this allows all mass parameters in the 5-d theory, including
$k$ and $1/r_c$ to lie in the vicinity of the Planck scale, the solution can
be deemed natural.

The above set-up, though particularly successful in resolving the
hierarchy issue without bringing in arbitrary intermediate energy
scales, has some unsatisfactory but inevitable features;
\begin{itemize}
\item The tension of the visible brane, fine tuned to realise 4-d
  Poincare invariance, turns out to be negative
  \cite{genrs,gothers}. This makes the brane configuration
  intrinsically unstable.
\item The visible brane, being flat, has zero cosmological constant.
  It is desirable to generalise the RS scenario to accommodate 
  non-zero values of this constant.
\item Though all the standard model (SM) particles were {\it assumed}
  to be confined to the visible brane in the original RS-theory
  \cite{rs}, later attempts revealed that all SM fermions and gauge
  bosons can, in principle, propagate in the fifth dimension
  \cite{bulkg, higgs-others1, bulkf1, bulkf}. The Higgs however
  remains as the only exception in the sense that the hierarchy issue
  with a bulk Higgs field cannot be addressed unless the ``minimal''
  set-up is extended with additional symmetries \cite{higgs-others1,
    bulkf1, chinese, higgs-others2, ssg1, cust}.
\end{itemize}

A recent work \cite{genrs} has suggested a solution to the first of
these problems by showing that one can take other solutions to the
warped geometry, where positive brane tension can be achieved. This
requires a negative bulk cosmological constant, as in the original RS
theory \cite{rs}, but also generates a non-zero cosmological constant
on the visible brane \cite{earlygen}. The modified warp factor of this
`generalised' RS scenario is a function of the induced cosmological
constant on the brane, and tends to the original RS exponential warp
factor in the limit when the brane cosmological constant goes to
zero. By demanding that the numerical value of the exponent has to be
the same as that in the minimal RS model, this warp factor produces
the required Planck-to TeV-scale warping, such that different values
of the induced brane cosmological constant correspond to different
values of curvature $k$, for a given value of the brane separation
scale $r_c$ (unlike the original RS model, where $kr_c~\simeq~12$
\cite{rs}).

Although the sign of the brane cosmological constant turns out be
negative in the region of the parameter space corresponding to
positive brane tension \cite{genrs}, its magnitude remains restricted
to rather small values, once this scenario is used to also generate
masses for neutrinos \cite{bmssg} via massive bulk neutrinos. The
observed small positive value of the cosmological constant still
necessitates some new physics on the 3-brane to cancel the induced
negative cosmological constant. However, the smaller magnitude lessens
the demand on such a cancellation making this scenario more
favourable. The stability of the visible brane makes it even more
tempting to studying particle phenomenology.

The object of this paper is to study whether this ``generalised'' RS
scenario can accommodate a bulk Higgs field while addressing the
hierarchy issue. In general, a number of difficulties in doing this
have been pointed out in the literature \cite{higgs-others1, bulkf1,
  chinese, higgs-others2}. These include

\begin{itemize}
\item The bulk Higgs vev, generated through spontaneous symmetry breaking
  (SSB) in bulk, is of large magnitude. This, in turn, lends a large bulk mass
  to the gauge bosons coupling to the Higgs in the bulk. Consequently, the
  lowest state in the KK tower of the gauge boson on the visible brane is
  inadmissibly massive, failing to comply with the mass requirement of $\le
  100$ GeV in the standard electroweak theory.

\item In the usual RS scenario, it has been shown that, for vanishing
  bulk mass of a bulk gauge boson, the first excited state in the KK
  tower acquires an unacceptably large coupling with matter. This puts
  a rather stringent restriction on the mass of such a state, in view
  of not only the direct search bounds at the Fermilab Tevatron but
  also in terms of precision electroweak constraints and limits on
  effective four-fermion interactions. It is possible to avoid such
  constraints if there is a bulk mass, as we shall show in section
  \ref{secfour}. However, there is still the need of explaining the
  case of the photon which cannot acquire a bulk mass, and is
  therefore beset with the problem related to its first massive
  excitation.
\end{itemize}

In this work, we suggest solutions to some of the above problems, and
indicate possibilities of solving the others, in the context of the
generalised RS scenario allowing a brane cosmological constant. We
emphasise at the outset that, while the essence of the RS philosophy
is to have bulk mass parameters only in the neighbourhood of the
Planck scale, this does not rule out $k$ and $1/r_c$ from being
somewhat less than, but close to, this order \cite{goldwise}. It is
thus our view that having both of the above parameters in the region
$10^{16-17}$ GeV is consistent with the spirit of an RS-like
theory. We show that, allowing simultaneously lowered values of $k$
and $1/r_c$, one obtains smaller values of the gauge boson mass on the
brane, thus fitting experimental data. In addition, we are also able
to show that the ratio between the gauge couplings of the lowest and
first excited states on the brane is brought down to a level where
direct search bounds from colliders are avoided.
 
In principle, this flexibility can be employed in the minimal RS model
\cite{rs} itself, by lowering the values of $k$ and $1/r_c$ in a
correlated manner. In spite of recognising this, we use the
generalised model \cite{genrs} to illustrate our point, for the
following reasons.  First of all, this model allows a positive brane
tension, thereby allowing a stable scenario. Secondly, it has been
employed earlier to accommodate a scheme of neutrino mass generation
\cite{bmssg}, where specific ranges of $k$ and $1/r_c$ are found to be
favoured by available data. We wish to see if a solution to the
problem of bulk Higgs mechanism can be found with parameters in the
same range. Thirdly, some solutions to the problem of the bulk 
photon scenario can be thought of, for example, by proposing part
of the symmetry breaking process to take place on the brane itself, by
utilising terms in the Higgs Lagrangian with a $\delta$-function
peaking at the visible brane. The success of such schemes often
depends on the existence of a curvature-Higgs coupling term in the
bulk along with a higher curvature Gauss-Bonnet extension
\cite{ssg1}. Such a model however encounters a stability problem
\cite{ssg2} when the brane tension is negative. It is thus expected
that the generalised RS scheme will accommodate such terms, whereby
the issue of symmetry breaking can be partly transferred to the brane,
thus making a massless photon (and some other features of W-and
Z-interactions) realistic. We maintain that our main thrust is on
suggesting the scheme of the Higgs vev in the bulk generating
acceptable massive gauge bosons on the brane. Indeed, there are
difficulties with various precision observables in spite of this,
which may require additional new physics. However, that does not
undermine the basic scheme proposed here, and the usefulness of having
the chosen generalised RS scenario.

We organise our paper as follows. In Section 2, we outline the main problems
of keeping the Higgs in the 5-d bulk if one sticks to the original RS
proposition. In Section 3, we briefly describe the essential features of the
generalised RS scenario. The possibility of having the Higgs in the bulk in
this scenario, and some numerical results of our study, are presented in
section 4. We summarise and conclude in section 5.

\section{Problems with a bulk Higgs field}
\label{sectwo}
The problem of having the mechanism of spontaneous symmetry breaking
in the 5-d bulk in the original RS formalism has already been
discussed in detail in several works \cite{higgs-others1, bulkf1,
  chinese, higgs-others2}. In this section we briefly outline the
essential points by assuming that the gauge bosons and the Higgs are
bulk fields. The fermions can either be bulk fields, or
brane-localised.

If the Higgs boson ($H$) is assumed to be a bulk field, the
corresponding 5-d potential takes the form,
\begin{equation}
\label{pot}
V(H) = -\mu^2 H^{\dagger}H + \frac{\lambda_{5d}}{2} (H^{\dagger}H)^2
\end{equation}
where $\mu>0$, and the Higgs field develops a vacuum expectation value
(vev) in the bulk $\sim \sqrt{\mu^2/\lambda_{5d}}$. $H$ is weakly coupled
when $<H>$, its vev, is of the same order as $\mu$.
This vev will
generate a bulk mass term $M$ for the bulk gauge boson $A_s [\equiv
A_\rho, A_5] (x_\mu, \phi)$, which, being a 5-d parameter, should be
of the order of Planck scale to be consistent with the basic spirit of
the RS scenario \cite{rs}. Through a process similar to that in 4-d, 
$M\sim <H>$.

Starting from the 5-d Lagrangian for the
gauge fields in the RS background and making use of $Z_2$ orbifold
conditions $\partial_5 A_\mu (x_\mu, \phi=\phi_i) = 0 = A_5 (x_\mu,
\phi=\phi_i)$ on the boundaries $\phi_i$, one can arrive at the
equation,
\begin{equation}
\label{orgrseq} 
\partial ^\mu (\partial _\mu A^a_\rho) - \frac{1}{r^2_c} \frac{\partial} 
{\partial \phi} \left( e^{-2\sigma}\frac{\partial}{\partial\phi}A^a_\rho \right)
+ M^2 e^{-2\sigma} A^a_\rho = 0 
\end{equation}
in the $A^a_5 =0$ gauge, $a$ being the gauge index. 
The 5-d field $A^a_\rho$ can be expanded into 4-d
KK modes as,
\begin{equation}
\label{orgrseq1} 
A^a_\rho (x_\mu, \phi) = \sum_n A^{a(n)}_\rho (x_\mu) 
\frac{f_n (\phi)}{\sqrt{r_c}}
\end{equation}
where each KK mode satisfies $(\Box + m^2_n) A^{a(n)}_\rho = 0$, $m_n$
being the corresponding KK mass. 

It should be noted that equation (2.2) is valid for both Abelian and
non-Abelian gauge fields that acquire mass through the vev of $H$
in 5-dimensions. This is because each component of $A^a_\mu$ still
satisfies   $(\Box + m^2_n) A^{a(n)}_\rho = 0$ \cite{peccei}, when one
considers the `free' part of its lagrangian, leaving out the self-interactions,
a procedure that is routine when one expands the field in normal modes. Thus 
the solutions of (2.2) should be the starting point in the KK decomposition of
non-Abelian gauge fields as well.

Using this expansion in
Eq.(\ref{orgrseq}) we get the relation satisfied by $f_n$ as
\begin{equation}
\label{orgrseq2} 
-\frac{1}{r^2_c} \frac{\partial}{\partial\phi} \left( e^{-2\sigma} 
  \frac{\partial}{\partial\phi} f_n \right) + M^2 e^{-2\sigma} f_n = m^2_n f_n
\end{equation}
The transformations $f_n = e^\sigma \chi_n$ and $Z_n = e^\sigma m_n /
k$ modify the above equation into the much known form
\begin{equation}
\label{orgrseq3} 
Z^2_n \frac{d^2\chi_n}{dZ^2_n} + Z_n \frac{d\chi_n}{dZ_n} + \left[
  Z^2_n - \left(1+\frac{M^2}{k^2}\right) \right] \chi_n = 0
\end{equation}
the solutions of which are the Bessel functions $J_{\alpha}$ of order
$\alpha = \sqrt{(1+M^2/k^2)}$. Thus one arrives at 
\begin{equation}
\label{orgrseq4} 
f_n = \frac{e^\sigma}{N_n} J_\alpha (\frac {m_n}{k}e^\sigma )
\end{equation}
where $N_n$ normalises $f_n$. The corresponding mass eigenvalues $m_n$
can then be found by solving the eigenvalue equation,
\begin{equation}
\label{orgrseig}  
\frac{x_n}{2} J_{\alpha-1} (x_n) + J_{\alpha} (x_n) - \frac{x_n}{2} 
J_{\alpha+1} (x_n) = 0, 
\end{equation}
which is obtained from the continuity of $df_n/d\phi$ on the visible
brane ($\phi=\pi$). Here
\begin{equation}
\label{orgrseig1}
x_n = m_n e^{\sigma(\pi)}/k
\end{equation}
The smallest root $x_0$ of the eigenvalue equation yields the lowest
mass eigenvalue $m_0$ and, the corresponding KK mode $A^{(0)}_\rho
(x_\mu)$ (the so called ``zeroth'' KK state) is interpreted as the
standard model (SM) field.
\begin{table}
\begin{center}
\begin{tabular}{|c|c|c|c|c|c|c|c|c|}
  \hline
  & \multicolumn{2}{|c|}{$M=0$} & \multicolumn{2}{|c|}{$M=0.1k$} 
& \multicolumn{2}{|c|}{$M=0.5k$} &  \multicolumn{2}{|c|}{$M=k$} \\ 
\hline
$n$ & $x_n$ & $m_n$ & $x_n$ & $m_n$ & $x_n$ & $m_n$ & $x_n$ & $m_n$ \\ 
\hline
0 & 0 & 0 & 2.41 & 2.41 & 2.54 & 2.54 & 2.87 & 2.87 \\
1 & 2.40 & 2.40 & 5.53 & 5.53 & 5.68 & 5.68 & 6.09 & 6.09 \\
2 & 5.52 & 5.52 & 8.66 & 8.66 & 8.82 & 8.82 & 9.25 & 9.25 \\ 
3 & 8.65 & 8.65& 11.80 & 11.80 & 11.97 & 11.97 & 12.40 & 12.40 \\
\hline
\end{tabular}\\
\caption {\small \it {Solutions of the eigenvalue equation for
    obtaining masses of KK gauge bosons in the RS-model for different
    ratios of $M/k$ are listed. For $k\sim M_P$, the corresponding
    masses are given in TeV.}}
\label{tab:1}       
\end{center}
\end{table}
If we solve Eq.(\ref{orgrseig}) for $M=0$, the lowest eigenvalue turns
out to be $m_0=0$ (see Table \ref{tab:1}). Following the previous
argument therefore, we know that this has to be the case for massless
SM gauge bosons like photon \cite{bulkf1, chinese}. Similarly, for the
massive ones like $W$ or $Z$, we should start with a non-zero $M$
($\sim M_P$), and expect $m_0 \simeq 100$ GeV. However, when we
actually solve Eq.(\ref{orgrseig}) for this case, the lowest {\it
  non-trivial} solution is obtained at $x_0 \sim 2$ for $M\sim k$
(note, from Eq.(\ref{orgrseq2}), that the zero-eigenvalue solution
demands the corresponding wave-function to vanish altogether and is
thus a trivial solution). In Table \ref{tab:1} we list the exact
values of $x_0$ corresponding to different non-zero ratios of
$M/k$. Recall that $e^{\sigma(\pi)}$ must be $\sim 10^{16}$ in order
to address the gauge hierarchy issue. As a result, from
Eq.(\ref{orgrseig1}) the lowest eigenvalue for this case turns out to
be $\sim 2$ TeV (see Table \ref{tab:1}). Naturally therefore, the
corresponding zeroth KK state in this case cannot be interpreted as
any of the known massive SM gauge bosons as such. Such an
interpretation will be possible if we could somehow lower the value of
$m_0$ to the required mass regime, which, in turn, can be achieved if
only we allow $M/k << 1$. This should be obvious from the fact that
the limit $M\to 0$ must reproduce $m_0 = 0$. In fact, it has been
shown in \cite{chinese} that to have $m_0$ in the 100 GeV regime, we
must have $M/k$ even lower than $\sim 10^{-10}$. This however is
undesirable, since this destroys the main spirit of the RS scenario by
bringing in a bulk mass parameter much smaller than the Planck mass.

It is, however, obvious from Eq.(\ref{orgrseig1}), that the above problem of
having the mass of the zeroth KK gauge boson in the required regime can be
addressed if one is allowed to lower the value of $k$ below the Planck
scale. But, the original formalism of RS does not allow such a freedom of $k$,
since, to address the hierarchy issue one must keep $kr_c \sim 12$
\cite{rs}. This is exactly where the generalised RS scenario may turn out to
be particularly useful. As mentioned before, the required warping for this
scenario does allow different values of curvature $k$, for a given value of
the brane separation scale $r_c$ and the induced cosmological constant on the
visible brane \cite{genrs}. Thus, it provides a chance to tune the values of
the gauge field mass parameters. Furthermore it may be possible that those
relevant $k$ values correspond to positive tension for the visible brane. In
such a situation therefore one can successfully achieve a five dimensional
spontaneous symmetry breaking mechanism generating masses for SM fields on a
stable visible brane with non zero brane cosmological constant. To this end we
probe the generalised RS scenario in the next two sections.

\section{RS scenario with a generalised warp factor}
\label{secthree}
The details of a generalised RS scenario with a nonzero brane
cosmological constant has been discussed in detail in \cite{genrs, jd,
  jd1}. In this section we briefly talk about the essential features
of this scenario.

Instead of the metric of Eq.(\ref{metric}) suppose we use one with a
more general warp factor $A(y)$,
\be 
\label{genmetric}
ds^2 = e^{-2A (y)} \eta_{\mu\nu} dx^\mu dx^\nu - dy^2 
\ee 
and evaluate $A(y)$ by extremising the action. As \cite{genrs} shows,
starting with an anti-de Sitter (AdS) bulk ($\Lambda < 0$), it is
possible to have a constant curvature brane spacetime which can be
either AdS ($\Omega < 0$) or dS ($\Omega > 0$) where $\Omega$ is the
induced cosmological constant on the visible brane. 

For $\Omega < 0$, the solution for the warp factor turns out to be
\be 
\label{genwarpf}
e^{-A(y)}          =     \omega         \cosh      \left(       \ln
  \frac{\omega}{1+\sqrt{1-\omega^2}} + ky \right)
\ee
where $\omega^2 ~=~ -\Omega^2/k^2$ is the absolute value of the
dimensionless quantity obtained out of $\Omega$. Note that the
original RS solution $A=ky$ is recovered in the limit $\omega^2 \to
0$. If we set $e^{-A}~\simeq~10^{-16}$ to ensure the hierarchy between
the Planck and EW scales, we find two solutions for $kr_c\pi$ for
every $\omega^2$, corresponding to positive and negative brane
tension, respectively. No solution, however, exists for $\omega^2 >
10^{-32}$. These solutions can then be used to determine the
corresponding visible brane tensions using the equation \cite{genrs}:
\begin{equation}
\label{gentension}
{\cal V}_{\rm vis} = 12M^3k \left[ \frac{\frac{\omega^2}{c_1^2} 
e^{2kr_c\pi}-1}{\frac{\omega^2}{c_1^2} e^{2kr_c\pi}+1}  \right ]  
\end{equation}
Recall that we are interested in having a positive
tension for the visible brane in order to have stable brane
configuration. One of the solutions, which yields the usual RS value
of $kr_c\pi~\simeq 36.84$ in the limit of near-vanishing $\omega$,
always corresponds to a negative brane tension, and hence is not
relevant to our cause. The other solution, on the other hand, leads to
positive brane tension and gives increasing values of $kr_c\pi$ as
$\omega^2$ decreases, leading to $kr_c\pi \simeq 250.07$ for $\omega^2
\rightarrow 10^{-124}$ (see Figures 1 and 2 in reference
\cite{genrs}). On the whole, a rather wide region in the $\omega^2 -
kr_c\pi$ space is generally allowed, a large part of which yields
positive brane tension (region III in Figure 1 in reference
\cite{genrs}). For $\Omega > 0$, it is not possible to have positive
tension for the visible brane, and hence, we exclude this case from
further discussions. 

Thus, the region of our interest in the $\omega^2 - kr_c\pi$ has a
negative brane cosmological constant, which is a necessary condition
for positive brane tension \cite{genrs,jd,jd1}. While the requirement
of a positive tension brane can lead to values of $\omega^2$ as high
as $10^{-32}$ (see Figure 1 in reference \cite{genrs}), it may be
desirable to keep $\omega^2$ as small as possible. This way one can
demand much less cancellation from some hitherto unknown physics to
play a role in its observed small positive value. Interestingly, if a
mechanism of neutrino mass generation is envisioned in terms of bulk
sterile neutrinos \cite{bulkf1,bulkf,gross,fermloc} in such a
generalised warped geometry, much smaller values of $\omega^2$ than
$10^{-32}$ can be shown to be consistent with phenomenology
\cite{bmssg}.

\section{Bulk Higgs in the generalised RS scenario}
\label{secfour}
If the Higgs is a bulk field in the generalised RS scenario described in
section \ref{secthree}, then proceeding just as in section \ref{sectwo}, the
equation of motion for the $n^{\rm th}$ KK mode ($f_n$) of mass $m_n$ of a
bulk gauge field is obtained as,
\bea 
\label{geneqn}
Z^2_n \frac{d^2\chi_n}{dZ^2_n} + Z_n \frac{d\chi_n}{dZ_n} + \left[
  Z^2_n - \left(1+\frac{M^2}{k^2}\right) \right] \chi_n = \nonumber \\
{\rm cosech}^2 (B) \left( Z_n \frac{d\chi_n}{dZ_n} - \left[ Z^2_n -
    \left(1+\frac{M^2}{k^2}\right) \right] \chi_n \right) 
\eea
where $Z_n = \frac{m_n}{k} e^A$, $\chi_n=e^{-A}f_n$, $c_1 =
1+\sqrt{1-\omega^2}$ and ${\rm cosech}^2 (B) = {\rm cosech}^2 ({\rm
  ln}( \omega/c_1)+kr_c\phi)$. It can be easily checked from
Eq.(\ref{genwarpf}) that the terms on the right hand side of the
above equation are proportional to $\omega^2$ or its higher
powers. Thus they vanish in the limit $\omega \to 0$, and as expected,
we get back the corresponding equation for the original RS model
\cite{rs}. In fact, as mentioned previously we will also be interested
in as small values of $\omega$ as possible \cite{bmssg}, so that we
can drop the right hand side of Eq.(\ref{geneqn}) for all practical
purposes. Thus, the solutions of $\chi_n$ from Eq.(\ref{geneqn}) will
still be the Bessel functions $J_{\alpha} (Z_n)$, and the eigenvalues
will, as usual, be obtained from the continuity of $f_n$ at
$\phi=\pi$,
\begin{equation}
\label{genrseig}  
\frac{x_n}{2} J_{\alpha-1} (x_n) + J_{\alpha} (x_n) - \frac{x_n}{2} 
J_{\alpha+1} (x_n) = 0 
\end{equation}
with $x_n$ now modified as \cite{jd1}
\begin{equation}
\label{genrseig1}
x_n = m_n e^{A(r_c\pi)}/k 
\end{equation}
and $\alpha = \sqrt{(1+M^2/k^2)}$ as before. This clearly means that
the solutions $x_n$ for different $n$ as listed in Table \ref{tab:1}
(for $M\ne 0$) are still valid (also listed in Table
\ref{tab:2}). Therefore, $x_0 \sim 2$ just as before, and we will have
to use $e^{A(r_c\pi)} \sim 10^{16}$ while evaluating the eigen value
$m_0$. However, since in this case we can indeed allow a somewhat
lower value for $k$, it is possible to have $m_0$ in the 100 GeV
regime; for example, $k \sim 4\times 10^{17}$ GeV yields $m_0 \sim 96$
GeV. For this choice of $k$, if, for example, for $\omega^2$ lying
in the range  $\sim 10^{-80} - 10^{-50}$, $1/r_c$ occurs around
$10^{16}$ GeV (see equation 21 in \cite{genrs}). This allows us to
simultaneously satisfy the neutrino data \cite{bmssg}. 
It is also clear from these figures 
that for smaller values of $\omega^2$, the
separation of the scales of $k$ and $1/r_c$ increases. In any case,
the zeroth KK mode in this scenario can indeed be identified with
known SM gauge bosons.
\begin{table}
\begin{center}
\begin{tabular}{|c|c|c|c|c|c|c|c|c|}
  \hline
  & \multicolumn{2}{|c|}{$M=0$} & \multicolumn{2}{|c|}{$M=0.1k$} 
& \multicolumn{2}{|c|}{$M=0.5k$} &  \multicolumn{2}{|c|}{$M=k$} \\ 
\hline
$n$ & $x_n$ & $m_n$ & $x_n$ & $m_n$ & $x_n$ & $m_n$ & $x_n$ & $m_n$ \\ 
\hline
0 & 0 & 0 & 2.41 & 96.40 & 2.54 & 101.60 & 2.87 & 114.80 \\
1 & 2.40 & 96.00 & 5.53 & 221.20 & 5.68 & 227.20 & 6.09 & 243.6 \\
2 & 5.52 & 220.80 & 8.66 & 346.40 & 8.82 & 352.8 & 9.25 & 370.00 \\ 
3 & 8.65 & 346.00 & 11.80 & 472.00 & 11.97 & 478.80 & 12.40 & 496.00 \\
\hline
\end{tabular}\\
\caption {\small \it {Solutions of the eigenvalue equation for
    obtaining masses of KK gauge bosons in the generalised RS-model
    for different ratios of $M/k$ are listed. Note that the solutions
    are exactly same as those obtained for the usual RS-theory (listed
    in Table I). For $k\sim 4\times 10^{17}$ GeV, the corresponding
    masses are given in GeV.}}
\label{tab:2}       
\end{center}
\end{table}

However, note from Eq.(\ref{genrseig1}), that such a lowering of the
value of $k$ not only lowers the mass of the zeroth KK state, it
lowers the masses of all the higher KK modes also. As a result, in
this generalised RS model with $k \sim 4\times 10^{17}$ the first KK
excitation corresponding to $x_1\sim 5.5$ has a mass $m_1 \sim 220$
GeV (see Table \ref{tab:2}). Recall that in the original RS-theory
with $K\sim M_P$, the same state had a mass $m_1 \sim 5.5$ TeV (see
Table \ref{tab:1}). In a similar way, the mass of the second KK state
will now be $m_2 \sim 350$ GeV instead of $9$ TeV in the original
RS-model, and so on. Whether or not these KK-states with such low
masses can survive the experimental constraints from direct
observations at the Tevatron \cite{teva} is decided by the strengths
of their corresponding couplings to SM fermions.

To estimate the strengths of such couplings, let us first assume that
the fermions are also bulk fields in this generalised RS scenario. We
refer the reader to Ref.\cite{jd} for a treatment of such
fermions. The 5-d Lagrangian describing the gauge interaction of the
bulk fermions with the bulk gauge field is then given as
\cite{chinese}
\begin{equation}
\label{orgrslag1}
e^{-1} {\cal L} = g_{5d} \bar \Psi (x^\mu,\phi) i\Gamma ^{\underline{S}}
e_{\underline{S}} ^{S}(\phi) A_S (x^\mu,\phi) \Psi (x^\mu,\phi)
\end{equation}
Using the expansions of both types of bulk fields in terms of KK excitations
in the above equation and integrating over the extra dimension, the expression
for the coupling of the $n^{\rm th}$ KK gauge boson to the massless zero-mode
fermion (i.e. SM fermion) bilinear is obtained as,
\begin{equation}
\label{orgrscoup}
g_n \sim g_{5d} \frac{\sqrt{2k}}{N_n (B_c - 1)} \frac{k^2}{m^2_n} 
\int^{1}_0  x dx[J_{\alpha} (x)]
\end{equation}
where $B_c = e^{A(r_c\pi)}$. Numerical evaluation of the above
integration yields $g_1/g_0 \sim 0.134$, $g_2/g_0 \sim 0.132$ and
$g_n/g_0 << 1$ for higher $n$. As a result of such reduced couplings,
the Tevatron limit of $M_T > 700$ GeV on the mass of a heavy vector
boson \cite{teva} (for a coupling of the same strength as in the SM) is
reduced to $M_T > 95$ GeV ($\sim 700 \times 0.134$). Thus, the KK
gauge bosons, although have somewhat lower masses for a low value of
$k$, still survive the experimental constraints.

Note that the ratios of the strengths of $g_n$ obtained in our case
have an interesting correspondence to those in section IV of
Ref.\cite{chinese} which deals with bulk gauge fields of zero bulk
mass.  Denoting their couplings by $\tilde g_n$, the correspondence
can be described as follows. The ratio $g_1/g_0$ in our case is $\sim
\tilde g_2/\tilde g_1$ in their case, and so on. In other words, the
SM gauge coupling $g_0$ in our case corresponds to the first KK gauge
coupling $\tilde g_1$ in their case, and so on. Such a correspondence
is easy to understand from the $x_n$ values listed in Table
\ref{tab:1} or \ref{tab:2} for both $M=0$ and $M \ne 0$. Note that
$x_0$ for $M \ne 0$ is $\sim x_1$ for $M=0$, $x_1$ for $M \ne 0$ is
$\sim x_2$ for $M=0$ and the like. This correspondence can then be
translated into a similar correspondence among the wavefunctions $f_n$
which are functions of $x_n$, and finally into the couplings ($g_n$ or
$\tilde g_n$) which are decided by these wavefunctions. Naturally, the
coupling $\tilde g_0$ in Ref.\cite{chinese} (i.e. the SM coupling in
their case) has no analogue in our case, since there is no zero-mass
gauge boson for $M \ne 0$.

It has been already mentioned that the next-excited state for any
spin-1 KK tower generally tends to have enhanced coupling, which may
not only cause conflicts with direct search bounds but also violate
the limits imposed by precision electroweak observables. For a gauge
boson which has acquired a bulk mass through SSB in five-dimensions,
this problem seems to be avoided, since this time one considers the
second massive mode in the tower vis-a-vis the first one, where the
former has relatively suppressed coupling with matter
\cite{chinese}. However, the problem may still persist for the KK
tower of the five-dimensional photon state which has no bulk mass.
Furthermore, one may encounter difficulties in complying with the
observed relations between the W-and Z-masses, or their relative gauge
couplings \cite{higgs-others1, chinese, higgs-others2}. As already
suggested, a number of remedies for this problem may be
considered. These include (a) some contact interaction on the brane
playing a role in the SSB of the U(1) symmetry, (b) a separate brane
kinetic energy term for the U(1) part of the SM \cite{ssg3}, and (c)
some altogether new physics (additional fields) which may cancel the
contributions to the precision electroweak variables.

Another interesting solution to the problem of enhanced coupling of
the first excited state of the photon can be envisioned as follows. If
one recalls Eq.({\ref{orgrscoup}}) for the coupling of the first
excited state, it becomes, on simplification,
\begin{equation}
\label{g1g}
g_1 \sim g  \frac{\sqrt{2\pi k r_c} k^2 e^{-2 A(r_c\pi)}}{m_1^2}
\end{equation}
In the generalised RS model the warp factor $A(r_c\pi)$ for $\omega^2 << 1$
is given (from equation 3.2) as
\begin{equation}
\label{g2g}
e^{-A(r_c\pi)} = e^{kr_c \pi} (\omega^2/4) + e^{-kr_c \pi}
\end{equation}
where the first term on the right-hand side dominates for values
of $kr_c$ corresponding to positive brane tension.

Unlike the standard RS model where a particular value of $kr_c$
uniquely determines the extent of warping on the SM brane \cite{rs},
here we have two-parameter solutions for the warp factor in terms of
$\omega^2$ and $kr_c$ \cite{genrs}. The two possible values of $kr_c$
for any given combination of $\omega$ and $e^{A(r_c\pi)}$ correspond
respectively to the positive and negative tension branes. If we wish
to focus on the branch of solution for which both the brane tensions
are positive (for the stability consideration) then the value of $kr_c
\pi$ must be greater than 37.5 \cite{genrs} i.e somewhat higher than
the usual RS value $\sim 36$ \cite{rs}.

In the region of solutions for $k r_c$ for which the brane tension is
positive we obtain from Eqs.(\ref{g1g}) and (\ref{g2g}),
\begin{equation}
\label{g3g}
g_1 \sim g  \frac{\sqrt{2\pi k r_c} k^2  (e^{kr_c \pi} (\omega^2/4))^2 }{m_1^2}
\end{equation}
In order to resolve the gauge hierarchy problem, the value of the warp
factor must be $10^{-16}$. This immediately yields different values of
$kr_c$ for different choices of the brane cosmological constant
$\omega^2$ from equation 4.7. If we now allow the value of the warp
factor to be $10^{-17}$ by appropriately choosing the value of the
cosmological constant $\omega^2$ one order lower than the value
required to make the warp factor $10^{-16}$, the coupling constant for
the first excited mode gets suppressed by one order (see
Eq.(\ref{g3g})). Thus at the expense of changing the warp factor by
just one order, we can remove the problem of unacceptable
enhancement of the coupling of the first KK mode of photon when the
SSB takes place in the bulk. Of course, in order to achieve acceptable
values of $m_n$ in Eq.(\ref{g2g}), one needs to have values of $k$
slightly different from that given in Table \ref{tab:2}. In fact in
this scheme the coupling of the first excited mode of the massive
vector boson also receives an additional suppression by an order (see
equation 4.5). Thus in the generalised RS model \cite{genrs} the
freedom of the choice of the brane cosmological constant and the
corresponding $kr_c$ provides us an alternative path to resolve this
long-standing problem of strongly coupled first excited mode for
massless gauge field in the bulk. This cannot be construed as
re-introducing the hierarchy problem, since an order of uncertainty in
the factor connecting the Planck and electroweak scales may arise due
to various kinds of unknown physics in the huge intervening range.

Before we end this section, it may be useful to comment on the issue
of fermion mass generation coming from five dimensional spontaneous
symmetry breaking with a bulk Higgs field. If the fermions ($F$) are
brane localised, the five dimensional Yukawa action will be given by,
\begin{eqnarray}
S^5_Y = \int d^4x \int dy [Y^5 \bar{F} F H \delta(y - \pi r_c)] 
\label{dig1}
\end{eqnarray}
where $Y^5$ is the five dimensional Yukawa coupling. When the symmetry
breaking takes place, the Higgs field should be replaced by $H = <H> +
h$, where $<H>$ is the vev developed in the bulk ($\sim
\sqrt{\mu^2/\lambda_{5d}} \sim$ Planck scale). The insertion of $<H>$
in the above equation generates masses for the fermions. The warp
factor in RS coming through the Jacobian leads to the suppression of
the large vev $<H>$ into a mass in the electroweak range
\cite{rs}. The observed spectrum of fermion masses, of course,
requires widespread values of $Y^5$, for which there is no explanation
as in SM. For the bulk fermion case, too, there exists a well-defined
prescription. In this case, all the KK fermions have two different sources
of mass, one being the ``Yukawa mass'' coming from SSB in the bulk,
and the other being the usual KK mass. The complete mass matrix for
this case is very similar to that described in \cite{huber}, although
the scenario itself is not the same as ours. The model in \cite{huber}
has bulk fermions and a brane localised Higgs field, while in our case
all are bulk fields. This only results in some non-zero non-diagonal
entries in the mass matrix which were zero in \cite{huber} (for
example, the (3,2) element of the matrix in equation 4.28 of that paper). 

Of course, when the fermions and the Higgs are both in the bulk, their
overlap can in general be large, thus making it difficult to explain
small fermion masses. This difficulty can be ameliorated by either
imposing some symmetry which makes the bulk Higgs profile peak near
the visible brane, or postulating a very small coefficient for the
Higgs-fermion coupling in the bulk. These are of course, tentative
proposals and are open to further investigations.

\section{Summary and conclusions}
We have studied bulk Higgs mechanism in an RS scenario where a cosmological
constant is induced in the brane. Such a scenario, as has been argued earlier,
ensures positive tension for both of the branes located at the orbifold fixed
points. Some mass parameters such as the induced bulk mass for vector bosons,
the curvature factor $k$ (related to the bulk cosmological constant) and the
inverse of the radius of the compact dimension are allowed to lie up to two
orders of magnitude below the Planck mass, a practice which, we argue, does
not go against the overall philosophy of RS-like theories.

We show that, once this is done, the KK tower of vector boson masses on the
visible brane becomes phenomenologically acceptable. First, the lowest-lying
member of the tower can have a mass within 100 GeV, thus raising hopes of
answering to features of the standard electroweak model. In addition, the next
excited state now is shown to have considerably reduced coupling to the
standard fermions, thus avoiding phenomenological constraints which would have
otherwise rendered the scenario unacceptable. The enhanced flexibility of
varying $k$ and $1/r_c$ enlarges the model parameter space consistent with
data. Needless to say, this is feasible in this scenario because of the
non-zero cosmological constant induced on the brane. Finally, we show that a
somewhat less rigid allowed range for the warp factor reduces the gauge
coupling of the first excited mode of the bulk photon, thereby allowing it to
have masses on a low scale.

\section{Acknowledgment}

PD is supported through the Gottfried Wilhelm Leibniz program by the
Deutsche Forschungsgemeinschaft (DFG). The work of BM was partially
supported by funding available from the Department of Atomic Energy,
Government of India, for the Regional Center for Accelerator-based
Particle Physics (RECAPP), Harish-Chandra Research Institute. SSG
acknowledges the hospitality of the Regional Centre for Accelerator-based
Particle Physics, Harish-Chandra Research Institute,
where part of the work was done.


\end{document}